\documentstyle[prl,floats,aps,twocolumn]{revtex}

\begin{document}
\draft
\title{The Entropy of the K--Satisfiability Problem}
\author{R\'emi Monasson \cite{rm} and Riccardo Zecchina \cite{rz}}
\address{\cite{rm} Laboratoire de Physique Th\'eorique de l'ENS,
24 rue Lhomond, 75231 Paris cedex 05, France\\ 
\cite{rz} INFN and Dip. di Fisica, Politecnico di Torino,
C.so Duca degli Abruzzi 24, I-10129 Torino, Italy}
\maketitle

\begin{abstract}

The threshold behaviour of the K--Satisfiability problem is studied in
the framework of the statistical mechanics of random diluted systems.
We find that at the transition the entropy is finite and hence that
the transition itself is due to the abrupt appearance of logical 
contradictions in all solutions and not to the progressive decreasing 
of the number of these solutions down to zero.
A physical interpretation is given for the
different cases $K=1$, $K=2$ and $K \geq 3$.

\end{abstract}


\narrowtext

The observation of critical behavior in randomly generated
combinatorial structures -- in mathematical and computer science
problems as well as in physical and biological models -- has recently
focused revived interest in the {\it Satisfiability} (SAT) problem for
randomly generated Boolean formulas as prototype of
NP--complete\cite{NPC} problems exhibiting threshold behaviour and
intractability concentration phenomena \cite{KirkSel,KirkII,Upper}.
The SAT problem is the root problem of complexity theory \cite{NPC}
and consists in determining the existence of an assignment of Boolean
variables $\{x_i=0,1\}_{i=1,\ldots,N}$ that evaluates to true a
generic conjunction (logical AND operation) of a set of clauses. Each
clause $\{C_\ell\}_{\ell=1,\ldots,M}$ is in turn defined as the
logical disjunction (logical OR operation) of a subset of literals,
which are either the variables $x_i$ or their negations $\bar x_i$.
The overall Boolean formula evaluates to true if and only if all
clauses are simultaneously satisfied, i.e. iff at least one among the
literals in each clause takes the Boolean value $1$.  K--SAT is a
version of SAT in which each clause contains a random set of exactly
$K$ literals.  When the number of clauses becomes of the same order as
the number of variables $(M=\alpha N)$ and in the large $N$ limit
(indeed the case of interest also in the fields of computer science
and artificial intelligence \cite{KirkII,Upper}), K--SAT exhibits
threshold phenomena. Numerical simulations show that the probability
of finding a correct Boolean assignment falls abruptly from one down
to zero when $\alpha$ crosses a critical value $\alpha_c(K)$ of the
number of clauses per variable. Above $\alpha_c(K)$, all clauses
cannot be satisfied any longer and one would rather minimize the
number of unsatisfiable clauses, which is the optimization version of
K--SAT also referred to as the MAX--K--SAT.

This scenario has been proven to be true in the $K=2$ case. The
mapping of 2--SAT on directed graph theory \cite{exact} indeed allows
to derive rigorously the threshold value $\alpha _c=1$ and an explicit
2--SAT polynomial algorithm working for $\alpha <\alpha _c$ has been
developed\cite{algo} (whereas, for $\alpha>1$, MAX--2--SAT is a
NP--complete \cite{Yannakakis,KirkII}).

For $K\ge3$, much less is known and not only MAX--K--SAT but also
K--SAT belongs to the NP--complete class \cite{NPC}. Some bounds on
$\alpha _c(K)$ have been derived\cite{Upper} and a remarkable
application of finite size scaling techniques has recently allowed to
find precise numerical values of $\alpha _c$ for $K=3,4,5,6$
\cite{KirkSel}. An important rigorous result is the self-averageness
taking place in MAX--K--SAT : independently of the particular sample
of $M$ clauses, the minimal fraction of violated clauses is narrowly
peaked around its mean value when $N\to\infty$ at fixed $\alpha$
\cite{SelfAve}. The situation becomes easier to understand in the
large $K$ limit where a simple probabilistic argument give the
asymptotic expression of $\alpha _c(K)\simeq 2^K \ln 2$\cite{KirkSel}.

The purpose of this letter is to study the K--SAT problem using
concepts and techniques of statistical mechanics of random diluted
systems.  To do so, we map K--SAT onto an energy--cost function by the
introduction of spin variables $S_i=\pm 1\,$ (a simple shift of the
Boolean variables) and of a quenched (unbiased) random matrix $C _{\ell ,i}=1$ 
(respectively $-1$) if $x_i$ (resp. $\bar x_i$) belongs to the clause
$C_\ell$, 0 otherwise. Then the function
\begin{equation}
E[C ,S]=\sum_{\ell =1}^M\,\delta \left[ \sum_{i=1}^N\,C _{\ell
,i}\,S_i \; ;\; -K\right] \; \; \;,
\label{energy}
\end{equation}
where $\delta [i;j]$ denotes the Kronecker symbol, 
equals the number of violated clauses and therefore its ground state
(GS) properties describe the
transition from K--SAT $(E_{GS}=0)$ to MAX--K--SAT $(E_{GS}>0)$.

While previous works on the statistical mechanics of other
combinatorial optimization problems - such as Traveling Salesman,
Graph Partitioning or Matching problems \cite{MPV,Optim} - focused
mainly on the study of the typical cost of optimal configurations, the
issues arising in K--SAT are of different nature.  Below $\alpha_c$,
the ground state energy vanishes and the key quantity to be analyzed
is the typical number of existing solutions, i.e. the ground state
entropy $S_{GS}$, for which no exact results are available so far.
Our main result is that $S_{GS}$ is still extensive at $\alpha=\alpha_c$~:
the transition is not due to
a progressive reduction of the number of solutions but to {\em the sudden
appearance of logical contradictions} in ``all'' of the exponentially
numerous solutions at the threshold.

In order to regularize the model, we compute the partition function
\begin{equation}
Z[C]=\sum_{\{S_i=\pm 1\}} \exp \left(-\beta E[C ,S] \right)
\label{Z} \; \; \;,
\end{equation}
after having introduced a finite ``temperature'' $1/\beta$. The
typical ground state free energy $\overline{ \ln Z[ C] } = \lim_{n \to
0} (\overline{Z[ C]^n}-1)/n$, where $\overline{(\dots)}$ stands for
the average over the random clauses, is then recovered in the limit of
zero temperature $\beta \to \infty$. The amount of technical work
necessary to perform the computation does not allow us to display all
details.  We then limit ourselves to a general description of the
methodological steps\cite{MPV,Optim} and focus mainly in the
discussion of the results. The complete calculation will be given in a
forthcoming paper\cite{MZlong}.

Once we have introduced $n$ replicas $S_i^a$, $a=1\ldots n$, of the
system, the average over the disorder $C$ couples all replicas
together through the overlaps $Q^{a_1,\dots ,a_{2r}}=\frac 1N \sum
_{i=1}^N S_i^{a_1}\ldots S_i^{a_{2r}}$ and their conjugated Lagrange
parameters $\hat Q^{a_1,\dots,a_{2r}}$ ($r=1,\dots,n/2$)\cite{Optim}.
The resulting effective Hamiltonian $N \;{\cal H} [\{Q\},\{\hat Q\}]$
involves all multi--replicas overlaps as expected in diluted
spin-glasses\cite{Optim,DeDom,gaussien} and is therefore much more
complicated than long--range disordered models where only interactions
between pairs of replicas appear\cite{MPV}. The free--energy is
evaluated by taking the saddle--point of ${\cal H}$ over all overlaps
$Q,\hat Q$.  This highly difficult task may be simplified by noticing
that, due to the indistinguishability of the $n$ replicas, the effective
Hamiltonian ${\cal H}$ must be invariant under any permutation of the
replicas. Therefore, one 
is allowed to look for a solution such that
the overlaps only depend upon the number of coupled replicas~:
$Q^{a_1,\dots ,a_{2r}}=Q_r$, $\hat Q^{a_1,\dots ,a_{2r}}=\hat
Q_r$\cite{MPV,Optim}. This is the so--called Replica Symmetric (RS)
Ansatz we shall use hereafter. Moreover, it results
convenient to characterize all $Q_r$ by introducing a probability
distribution $P(x)$ of the Boolean magnetization $x= \langle
H\rangle$, such that $Q_r=\int_{-1}^1dx P(x)x^{2r}$. Elimination of
the Lagrange parameters $\hat Q_r$'s leads to the expression
\begin{eqnarray}
\frac 1N\overline{\ln Z[C]} &=&\log 2 -\frac 12\int_{-1}^1dxP(x)\log 
(1-x^2) + \nonumber \\
&& \alpha (1-K)\int_{-1}^1
\prod_{\ell =1}^K dx_\ell P(x_\ell )\log A_{(K)} +  \nonumber \\
&&\frac{\alpha K}2\int_{-1}^1\prod_{\ell =1}^{K-1}dx_\ell P(x_\ell )\log
A_{(K-1)} \; \; \;, 
\label{Frs}
\end{eqnarray}
with $A _{(J)} \equiv A _{(J)} [\{ x_\ell \} ,\beta]= 
1+(e^{-\beta}-1)\prod_{\ell =1}^{J}(1+x_\ell)/2$ for $J=K-1$ and $J=K$.
The measure $P(x)$ is given by the saddle point integral equation
\begin{eqnarray}
P(x) &=&\frac 1{1-x^2}\int_{-\infty }^\infty du\cos \left[ \frac u2\log
\left( \frac{1+x}{1-x}\right) \right] \exp \bigg[ -\alpha K+
\nonumber \\ && \left. \alpha K\int_{-1}^1\prod_{\ell =1}^{K-1}dx_\ell
P(x_\ell )\cos \left( \frac u2\log A_{(K-1)} \right) \right] \; \; \;. 
\label{pxint}
\end{eqnarray}

%%%%%%%%%%%%%%%%%%%%%%%%%%%%%%%%%%%%%%%%%%%%%%%%%%%%%%%%%%%%%%%%%%%%%%%%%%
% K=1
%%%%%%%%%%%%%%%%%%%%%%%%%%%%%%%%%%%%%%%%%%%%%%%%%%%%%%%%%%%%%%%%%%%%%%%%%%
A toy version of the K--SAT problem is obtained when $K=1$. As we shall see,
some interesting information may already be obtained from this almost trivial 
case. A sample of $M$ clauses is completely described by the set of integer 
numbers $\{ t_i, f_i\}$, $i=1,\dots,N$, where $t_i$ (respectively
$f_i$) is the number of clauses imposing that $S_i$ must be true (resp.
false). The partition function (\ref{Z}) corresponding to this sample
reads $Z[\{ t_i, f_i\}]=\prod _{i=1}^N (e^{-\beta t_i}
+e^{-\beta f_i})$. Averaging over the probability weight $M! / \prod
_{i=1}^N (t_i! f_i!)/ (2N)^M$ of the sample, we find
$\frac 1N\overline{\ln Z} = \ln 2 - \alpha \beta/2 +
\sum _{l=-\infty}^{\infty} e^{-\alpha} I_l (\alpha)\;
\ln (\cosh (\beta l/2))$
where $I_l$ denotes the $l^{th}$ modified Bessel function.
In the limit of zero temperature, the energy and the entropy of the
ground state read $E_{GS}(\alpha )=\alpha [ 1-e^{-\alpha } I_0(\alpha ) - 
e^{-\alpha } I_1(\alpha )]/2$ and
$S_{GS}(\alpha )= e^{-\alpha } I_0(\alpha )\; \ln 2$ respectively.
Therefore, as soon as $\alpha$ is non zero, the
clauses cannot be satisfied all together but there is an exponentially large
number of different values for the Boolean variables giving the same minimum
fraction $E_{GS}(\alpha )/\alpha$ of unsastisfiable clauses. 
The reason is that, though all Boolean variable are required to be
true ($t_i>0$) and false ($f_i>0$) at the same time, a finite fraction
of them, $e^{-\alpha } I_0(\alpha )$, fulfill the condition 
$t_i=f_i$. The latter can therefore be chosen at our convenience, without
changing the ground state energy.
These results may be found back within our approach, showing the RS Ansatz is
exact for all $\beta$ and $\alpha$ when $K=1$.  
The saddle--point equation for $P(x)$  can be explicitly 
solved at any temperature $1/\beta$ and the solution read
\begin{equation}
P(x)= \sum _{\ell=-\infty}^{\infty} e^{-\alpha} I_\ell (\alpha) \; \delta
\left( x - \tanh \left( \frac{\beta \ell}{2} \right) \right) \; \; \;. 
\label{Saddlek=1}
\end{equation}
In the limit of physical interest $\beta \to \infty$, $P(x)$ reduces
to a sum of three Dirac peaks in $x=\pm 1$ and $0$ with weights
$(1-e^{-\alpha } I_0(\alpha ))/2$ and  $e^{-\alpha } I_0(\alpha )$
respectively. It clearly appears
that the finite value of the ground state entropy is due to the
presence of unfrozen spins, resulting from the mechanism exposed
above.  This is an important feature of the problem which
remains valid for any $K$.

%%%%%%%%%%%%%%%%%%%%%%%%%%%%%%%%%%%%%%%%%%%%%%%%%%%%%%%%%%%%%%%%%%%%%%%%%%
% K cresce ...
%%%%%%%%%%%%%%%%%%%%%%%%%%%%%%%%%%%%%%%%%%%%%%%%%%%%%%%%%%%%%%%%%%%%%%%%%%

Another relevant mechanism is the accumulation of magnetizations
$\overline{\langle H \rangle} =\pm (1-O(e^{-|z| \beta}))$, $z=O(1)$,
giving two Dirac peaks contributions to $P(x)$ in $x=\pm 1$ in the
zero temperature limit, as can be seen from (\ref{Saddlek=1}).  The
occurrence of such peaks means that a finite fraction of spins are
frozen and hence that a further increase of $\alpha$ beyond $\alpha_c$
would cause the appearance of unsatisfiable clauses.

This scenario remains valid for any $K$.  The fraction of violated
clauses at temperature $1/\beta$ can indeed be computed through
$E=-\frac 1N\partial \overline{\ln Z}/\partial \beta$.
The ground state energy will clearly depend only upon the
magnetizations of order $\pm (1-O(e^{-|z|\beta}))$, if any. These
contributions can be picked up by the new function 
$R(z)=\lim _{\beta \to \infty} \beta P(\tanh( \beta z/2)) /2/ \cosh^2(
\beta z/2)$ satisfying the saddle--point equation
\begin{eqnarray}
R(z)=\int_{-\infty}^{\infty}\frac{du}{2\pi}\cos(uz) \exp \bigg[
-\frac{\alpha K}{2^{K-1}} + \alpha K \times \nonumber \\
\left. \int _0 ^{\infty}\prod_{\ell=1}^{K-1} 
dz_\ell R(z_\ell)\cos( u\ \hbox{\rm min} (1,z_1,\ldots,z_{K-1}))\right] \; \;.
\label{Saddler}
\end{eqnarray}
Remarkably, it is possible to find analytically an exact solution to this
functional equation for any $K$ and $\alpha$~:
\begin{equation}
R(z)=\sum_{\ell=-\infty}^\infty e^{-\gamma} I_\ell(\gamma) \delta 
(z-\ell)\ \ ,
\end{equation}
where $\gamma$ is solution of the implicit equation
\begin{equation}
\gamma=\alpha K \left[\frac{1-e^{-\gamma} I_0 
(\gamma)}{2}\right] ^{K-1} \ \ .
\label{Miracolo}
\end{equation}
The corresponding cost function equals $E_{GS}(\alpha)=\gamma ( 1- 
e^{-\gamma} I_0(\gamma)- K e^{-\gamma} I_1(\gamma))/2/K$.
We shall now analyse the physical structure of this solution and show how the
predictions it leads to for $K=2$ qualitatively differ with respect to the
case $K\ge 3$.

%%%%%%%%%%%%%%%%%%%%%%%%%%%%%%%%%%%%%%%%%%%%%%%%%%%%%%%%%%%%%%%%%%%%%%%%%%
% K=2
%%%%%%%%%%%%%%%%%%%%%%%%%%%%%%%%%%%%%%%%%%%%%%%%%%%%%%%%%%%%%%%%%%%%%%%%%%

Self--consistency equation (\ref{Miracolo}) admits the solution
$\gamma=0$ for any $\alpha$ \cite{Falso}. When $K=2$, there is another
solution $\gamma(\alpha)>0$ above $\alpha=1$.  This new solution
maximizes $E_{GS}$ (and then the free--energy) and must be
preferred\cite{MPV}.  Therefore, our RS theory predicts that
$E_{GS}=0$ for $\alpha \le 1$ and increases continuously when $\alpha
>1$, giving back the rigorous result $\alpha_c(2)=1$. The transition
taking place at $\alpha_c$ is of second order with respect to the
order parameter~: the value of $\gamma$ does not show any jump and two
Dirac peaks for $P(x)$ progressively appears in $x=\pm 1$ with
amplitude $(1-e^{-\gamma}I_0(\gamma))/2$ each. For large $\alpha$, the
RS ground state energy scales as $E_{GS}\simeq \alpha/ 4$ which is
known to be exact\cite{KirkII}.  As far as 2--SAT is concerned, the
value of the threshold is correctly predicted and one can reasonably
assume that the RS solution is valid below $\alpha=\alpha_c=1$.  Above
$\alpha_c$, further analysis is needed \cite{MZlong} in order to
discuss the exactness of the RS solution.

For $\alpha > \alpha_c$, there do not exist anymore sets of $S_i$'s such that
the energy (\ref{energy}) remains non extensive.  The vanishing of the
exponentially large number of solutions that were present below the
threshold is surprisingly abrupt.  We have indeed studied the number
of such solutions as a function of the number of clauses per spin in
the range $0\le\alpha\le\alpha_c$. Their logarithm (divided by $N$),
that is the entropy of the ground state $S_{GS}(\alpha)$, is given by
(\ref{Frs}) when $\beta\to\infty$. Finding the solution $P(x)$ of the
implicit function equation (\ref{pxint}) is a difficult numerical
task. We have therefore resorted to an exact expansion of $P(x)$ in
powers of $\alpha$, starting from $P(x)|_{\alpha=0}=\delta(x)$, and
injected the resulting probability function into (\ref{Frs}) to obtain
the expansion of $S_{GS}(\alpha)$. 
At the $8$-th order (which implies an uncertainty less than one percent),
we have found that $S_{GS}(\alpha_c)\simeq 0.38$, which is
still very high as compared to $S_{GS}(0)=\ln 2$ (see fig. 1). It is remarkable
that the entropy does not vanish at the transition but keeps an
extensive value just below the threshold. The transition is therefore due to
the abrupt appearance of contradictory logical loops in ``all''
solutions at $\alpha=\alpha_c$ and not to the progressive decreasing of the
number of these solutions down to zero at the threshold.

%%%%%%%%%%%%%%%%%%%%%%%%%%%%%%%%%%%%%%%%%%%%%%%%%%%%%%%%%%%%%%%%%%%%%%%%%%
% K >= 3
%%%%%%%%%%%%%%%%%%%%%%%%%%%%%%%%%%%%%%%%%%%%%%%%%%%%%%%%%%%%%%%%%%%%%%%%%%

Let us turn now to the $K\ge 3$ case. Resolution of implicit equation
(\ref{Miracolo}) leads to the following picture. For $\alpha < \alpha
_m(K)$, there exists the solution $\gamma=0$ only. At $\alpha _m(K)$, a
non zero solution $\gamma (\alpha)$ discontinuously appears.  The
corresponding ground state energy is negative in the range
$\alpha_m(K) \le \alpha <\alpha_s(K)$, meaning that the new solution
is metastable and that $E_{GS}=0$ up to $\alpha_s(K)$. For
$\alpha>\alpha_s(K)$ the $\gamma (\alpha) \neq 0$ solution becomes
thermodynamically stable, leading to the conclusion that $\alpha_s(K)$
corresponds to the desired threshold $\alpha_c(K)$.  However, this
prediction is wrong as can be immediately seen for $K=3$, since the
experimental value $\alpha_c(3)=4.17\pm0.05$\cite{KirkSel} is lower
than $\alpha_m(3)\simeq 4.667$ and $\alpha_s(3)\simeq 5.181$.  In
addition, large $K$ evaluations give $\alpha _m(K)\sim K 2^K/16
/\pi$ and $\alpha_s(K)\sim K 2^K /4/\pi$, which grow faster than the
asymptotic value $\alpha_c(K)\sim 2^K \ln 2$\cite{energy}.

This situation, strongly reminiscent of neural networks with binary
couplings\cite{Binary}, may be understood by an inspection of the RS
ground state entropy. To do so, we have expanded $S_{GS}$ to the
$\ell^{th}$ order in $\alpha$ using the same method as for $K=2$ and
denoted by $\alpha_{ze}^{(\ell)}(K)$ the point where it vanishes.
Note that $\alpha_{ze}^{(1)}(K)$ corresponds to the annealed theory
while $\alpha_{ze}^{(\ell)}(K)$ converges to $\alpha_{ze}(K)$ when
$\ell\to \infty$, that is the exact value of $\alpha$ at which
$S_{GS}$ goes to zero. For $K=3$, we have performed the expansion up
to $\ell=8$ and found~: $\alpha_{ze}^{(1)}=5.1909$,
$\alpha_{ze}^{(2)}=5.0144$, $\alpha_{ze}^{(3)}=4.9189$,
$\alpha_{ze}^{(4)}=4.8589$, $\alpha_{ze}^{(5)}=4.8187$,
$\alpha_{ze}^{(6)}=4.7893$, $\alpha_{ze}^{(7)}=4.7677$,
$\alpha_{ze}^{(8)}=4.7504$, indicating that $\alpha_{ze}$ is
definitively larger that $\alpha_c\simeq 4.17$.  Repeating the
calculation for $K=4,5,6$, we have obtained qualitatively similar
results which show an even quicker convergence towards a zero entropy
point such that $\alpha_c(K)< \alpha_{ze}(K) < \alpha_s (K)$.
Finally, in the large $K$ limit, $\alpha_{ze}(K)$ asymptotically reaches the
threshold $\alpha_c(K)$ from above.  As $K$ grows, fluctuations get
weaker and weaker and Gaussian RS theory becomes exact\cite{gaussien}.
Solving equation (\ref{pxint}), we find for $K\gg 1$ and $\alpha \le
\alpha_c(K)$
\begin{equation}
P(x) \simeq \frac{1}{\sqrt{2 \pi u(\alpha)} (1-x^2)} \exp \left(
-\frac{1}{8 u(\alpha)} \log^2 \left( \frac{1+x}{1-x} \right) \right)
\label{pxasint}
\end{equation}
where $u(\alpha)=\alpha K/4^{K-1}$. As a consequence, $P(x)\to \delta(x)$ 
when $K\to\infty$ tells us that the annealed
theory becomes exact in the large $K$ limit.

Therefore, the situation is as follows for (finite) $K\ge 3$.  Above
$\alpha_{ze}$, the RS entropy is negative whereas it has to be the
logarithm of an integer number. The RS Ansatz is clearly unphysical in
this range, explaining why $\alpha_s$ and $\alpha_c$ do not coincide.
At the threshold $\alpha_c$ (which is experimentally known to be lower
than $\alpha_{ze}$), the RS entropy is still extensive.  The crucial
question now arises whether this result is exact or is affected by
Replica Symmetry Breaking (RSB) effects.
To clear up this dilemma, an analysis of RSB effects would
be required.  Due to the general complexity of such an approach in
diluted models\cite{DeDom} and the technical difficulty of the K--SAT
problem, the preliminary attempts we have done in this direction have
not been successful yet\cite{MZlong}.  We have then resorted to
exhaustive numerical simulation in the range $N=12,...,28$ and
compared the corresponding ground state entropies $S_{GS}^{(N)}
(\alpha)$ to our RS theory for K=3.  As reported in Fig.1, for
$\alpha<\alpha_c$, our analytical solution agrees very well with the
numerical results. This confirms that the entropy of the ground state
is finite at the threshold.  The comparison may be made more precise by
a careful extrapolation of the entropy $S_{GS}^{(N)} (\alpha \simeq
4.17)$ in $1/N$ (see inset of Fig.1). The extrapolated value
appears to be in perfect agreement with the RS
prediction $S_{GS} \simeq 0.1 $.  Therefore, RSB corrections to the RS
theory seem to be absent below $\alpha_c$, which leads us to think
that RSB could occur at $\alpha_c$ exactly.

To conclude, let us say that one should however not deduce from the
above remark that the structure of the solution--space is simple.  It
might well happen that the solution--space could have a non trivial
structure which is not reflected by the magnetization distribution
$P(x)$ only\cite{nascosto}. It would be very interesting to
understand if such a phenomenon could take place in the K--SAT problem
and what information the hidden structure of the solution--space could
give us about its algorithmic complexity.

\begin{figure}
%\epsfbox{fig.ps}
\caption{Entropy vs. $\alpha$ for $K=2$ and $3$. 
The analytical solutions (solid lines) are compared with numerical 
exhaustive simulations for $N=12, 16, 20, 24$ and 
$30000, 15000, 7500, 2500$
samples respectively (for $K=2$ we stop at $\alpha=2.5$ whereas for $K=3$
at $\alpha=6$). Error bars are within $10\%$ and thus not reported
explicitly. Inset: $1/N$ entropy extrapolation for $\alpha=4.17$
(in average for each $N$), $N=20,22,24,26,28$ 
(with $16500, 11500, 7500, 4000, 3000$ samples respectively) and $K=3$.} 
\label{fig1}
\end{figure}


\begin{thebibliography}{99}
\bibitem{Falso}
This is true if $K>1$. For $K=1$, the only solution is $\gamma=\alpha$ 
which gives back all results already derived in this case at zero temperature.
\bibitem{energy}
Though the scaling of $\alpha_c(K)$ for large $K$ is wrong within the
RS Ansatz, the asymptotic value for large $\alpha$ (and any $K$) of
the ground state energy for MAX--K--SAT is correctly predicted~:
$E_{GS}(\alpha) \sim \alpha / 2^K$ \cite{SelfAve}.
\bibitem{gaussien}
This is not surprising since $K$ may be interpreted as the
number of ``neighbors'' to which each spin is coupled inside a
clause.
\end{thebibliography}
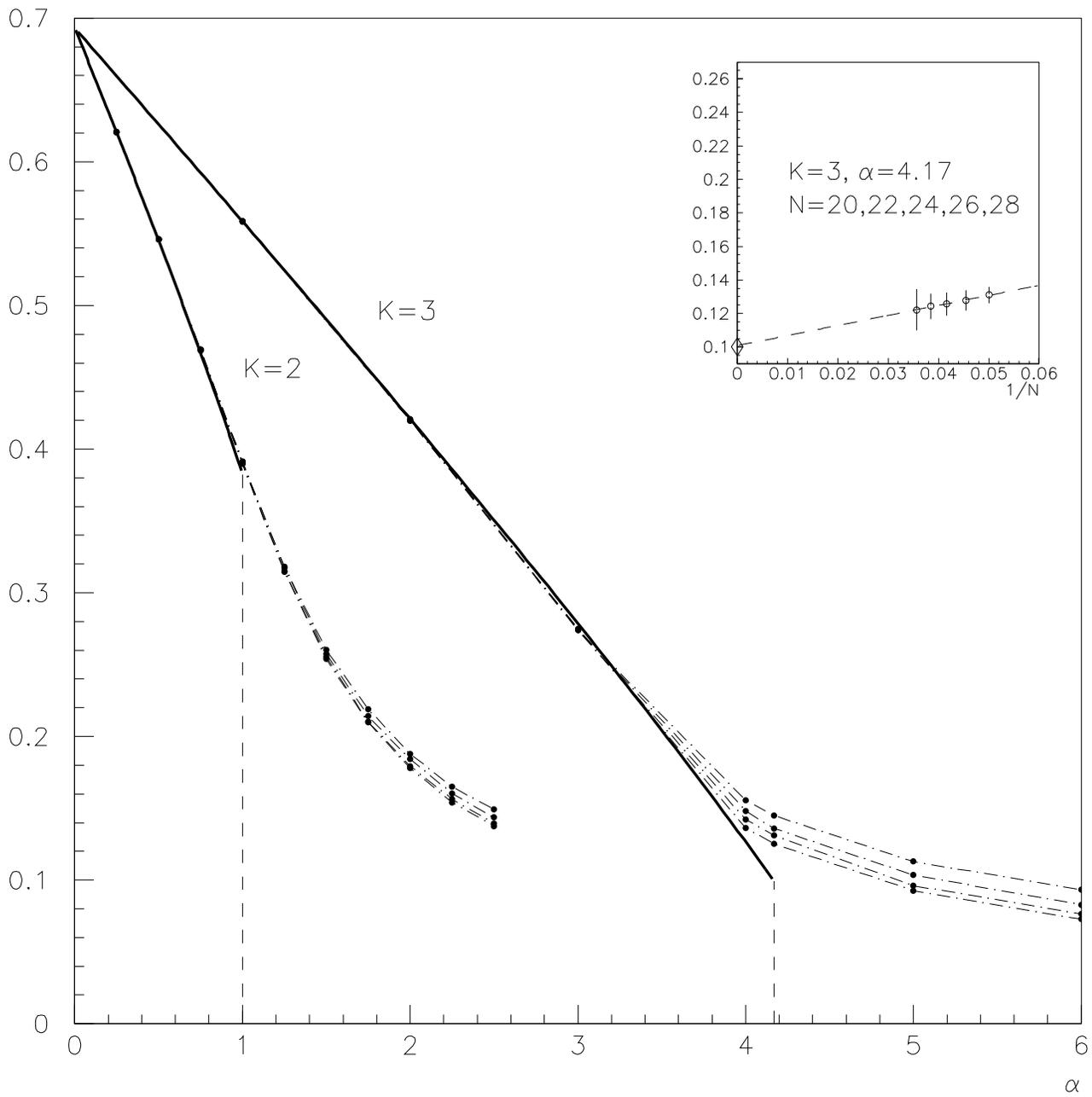
\end{document}